\documentclass{article}
\usepackage[utf8]{inputenc}
\usepackage{graphicx}
\usepackage{amsmath}
\usepackage{amssymb}
\usepackage{multirow}
\usepackage{lipsum}
\usepackage{verbatim}
\usepackage{xcolor}

\usepackage[a4paper, left=20mm, right=20mm, top= 30mm]{geometry}

\title{A direct approach to coherent states of billiards using a quantum algebra framework }
\author{A. C. Maioli\footnotemark[1], E. M. F. Curado\footnote{Centro Brasileiro de Pesquisas F\'{\i}sicas.\\ \textit{Rua Xavier Sigaud 150, 22290-180, Rio de Janeiro, RJ, Brazil}}}

\begin{document}

\maketitle

\begin{abstract}
    
Quantum billiards are a key focus in quantum mechanics, offering a simple yet powerful model to study complex quantum features. While the development of algebras for quantum systems is traced from one-dimensional integrable models to quantum groups and the Generalized Heisenberg Algebra (GHA).
The primary focus of this work is to extend the GHA to quantum billiards, showcasing its application to separable and non-separable billiards. We apply the formalism to a square billiard, first generating one-dimensional coherent states with specific quantum numbers and exploring their time evolution.Then, we extend this approach to develop two-dimensional coherent states for the square billiards. We also demonstrate its applicability in a non-separable equilateral triangle billiard, describing their algebra generators and associated one-dimensional coherent states. \\

\end{abstract}

%%%%%%%%%%%%%%%%%%%%%%%%%%%%%%%
\section{Introduction}
%Billiards

Quantum billiards have become an important research area in quantum mechanics as they provide a simple and powerful model for studying the behavior of complex quantum features. Such as universality laws in quantum chaology \cite{PhysRevE.107.L032201, RISER2021168393}, vibrating quantum systems \cite{PhysRevE.87.012912, doi:10.1063/1.166503}, experiments describing wave function morphologies \cite{ravnik2021quantum, cabosart2017recurrent}, eigenvalue obtainment via special functions or quantum scattering \cite{Heller1997, Maioli2018, Maioli2019, Nikulin_2024}, and so on \cite{stockmann2000quantum}. Furthermore, diverse quantum systems (quantum dots\cite{doi:10.1126/science.1154663, libisch2009graphene, nakamura2004quantum,nazmitdinov2012shape}, quantum rings \cite{hackens2006imaging}, resonance cavities \cite{Zanetti2012, Teston2022}, quantum corrals \cite{QuantumCorralHeller}, etc.) rely on quantum confinement, which can be depicted as quantum billiards, having many technological applications.
Another important aspect is the classical correspondence of quantum mechanics, which relates to exploring quantum states showing semi-classical time behavior. It started with Schrödinger's work on ``packets of eigenmodes" of the harmonic oscillator \cite{Schro}. According to Bohr's correspondence principle, as the quantum numbers of a system increase significantly, its behavior and predictions should approach those of classical physics. In the context of billiards, the usual wave functions do not seem to exhibit the characteristics of classical periodic orbits. Thereby, it is necessary for clever manipulations of the wave functions to approach this connection between the classical equations of motion (with closed orbits) and the energy spectrum \cite{gutzwiller1998resource,brack2018semiclassical,Bergeron2010,sym15101809}. Similarly, quantum revival is often studied because wave packets depend on the energy spectrum, and its time evolution disperses on various orbits. Then, the quantum-classical connection turns out to be evident through the calculations of the semi-classical quantum revival time \cite{Robinett2004}. In the context of quantum-classical correspondence, the important work of Heller (\cite{ PhysRevLett.53.1515, PhysRevA.46.R1728}, and references therein) show the association of the eigenfunctions with classical periodic orbits and scarred states.

%Coherent states
Coherent states are specific quantum states that exhibit classical-like properties while still being fully quantum mechanical.  Originally, the foremost one arose in Quantum Electrodynamics in 1953 \cite{PhysRev.91.728} while investigating the electromagnetic field, as perturbed by a prescribed current. Then, Glauber advances this topic (Coherent States) with his works in quantum optics \cite{PhysRevLett.10.84, PhysRev.130.2529, PhysRev.131.2766}.  Along these lines, coherent states are fundamental in studying quantum mechanics and quantum optics because they provide a bridge between classical and quantum descriptions of physical systems, as can be seen in the work of Sudarshan\cite{PhysRevLett.10.277}. Refer to pivotal articles such as \cite{RevModPhys.37.231, PhysRev.177.1857, PhysRevD.2.2161, PhysRevD.2.2187} for the subsequent advancements in quantum optics and quantum field theory. The quantum mechanical approach to coherent states was investigated by Klauder \cite{KLAUDER1960123, 10.1063/1.1704034, 10.1063/1.1704035}, for further reading about the properties see \cite{klauder1985coherent, perelomov1972coherent, AMPerelomov1977, Gazeaubook}.

%Quantum algebras
On the other side, the development of algebras aimed at quantum systems emerged in an approach to one-dimensional integrable quantum models using Bethe's algebraic ansatz \cite{bethe1, bethe2}. Then, the construction of different algebras was relevant to address a large class of physical phenomena. Some of them are related to two-dimensional solvable models for S-matrices and the solutions of their Yang-Baxter factorization equations \cite{bethe1, Sklyanin1, faddeev1988quantization, jimbo1988quantum}, Hamiltonian's of anisotropic spin chains \cite{pasquier1990common, batchelor1990q, kulish1991general}, topologic quantum field theory \cite{manin1989multiparametric}, Wilson loops in three-dimensional Chern-Simons theory \cite{witten1990gauge, guadagnini1990chern}, chiral vertices, fusion rules and blocks in conformal field theory \cite{alvarez1989quantum, gomez1990quantum}, two-dimensional Liouville gravity \cite{gervais1991quantum}, quantum statistics \cite{greenberg1991particles, fivel1990interpolation}, heuristic phenomenology of deformed molecules and nuclei \cite{iwao1990knot,raychev1990quantum}. When the algebra is associated with the harmonic oscillator, it is called Heisenberg algebra and is presented in terms of the creation and annihilation operators. Its generalization (q-oscillators) was implemented through the group $\textit{su}_q(2)$ in the Jordan-Schwinger method \cite{klimek1993extension}, which is a deformed version of the group $\textit{su}(2 )$.

%GHA
In this context, the Generalized Heisenberg algebra (GHA) was forged \cite{Curado2000, Curado2001, Curado2008} in early 2000. It consists of a Heisenberg-type algebra that relies on an arbitrary function $f$. This function depends on the dimensionless Hamiltonian and is carefully chosen to correspond to the desired quantum system, connecting the eigenvalues by means of the relation $\epsilon_{i+1} = f(\epsilon_i)$, where $\epsilon_i$ is the $i$-th energy eigenvalue of the system.
Therefore, it is possible to uncover ladder operators and construct coherent states for several quantum systems \cite{Hassouni2005}. The GHA was originally created to describe one-dimensional quantum systems \cite{Curado2001b}. In 2006 the GHA was implemented on a hydrogen atom \cite{Curado2006}, this is possible because of the eigenstate's degeneracy, in other words, the energy depends on only one of the quantum numbers. In this work, we employ the GHA to quantum systems where the energy depends on two quantum numbers; we construct the algebra related to one of the quantum numbers while the other is held fixed.

Aside from the generalized harmonic oscillator algebra (GHA) approach, there are various methods for generalizing the harmonic oscillator algebra, primarily developed to address the eigenvalue problem in different quantum systems without directly solving the associated differential equation. For example, the supersymmetric approach enables the determination of eigenvalues and eigenfunctions for nearly all analytically solvable one-dimensional potentials, including quantum systems with position-dependent effective mass. This approach has also been applied by several researchers to generalize coherent states (CS) within the framework of supersymmetric quantum mechanics. In this context, the work of Bergeron and Valance \cite{BergeronSupersymmetric} is noteworthy, as they used a supersymmetric formalism to derive oscillator-like coherent states for a one-dimensional Hamiltonian with an arbitrary potential.

%Objetivos
The main purpose of this work is to develop a generalized Heisenberg algebra for quantum billiards as it allows us to construct coherent states for these billiards in a much more direct way. 
Henceforth, we depict a subspace of Hamiltonian's eigenfunction via GHA for any quantum system with quadratic eigenenergies. We show a closed form of the normalization constant of the coherent state and the weight function, 
which allows a direct application of the GHA to obtain coherent states. We employ this formalism to a square billiard and find the algebra generators and the position representation of the ladder operators. Along these lines, we construct coherent states related to one of the quantum numbers, namely ``one-dimensional coherent states''. We investigate the time evolution of those states and within the quantum revival time. Then, we construct general ``two-dimensional coherent states" and their respective time evolution. By employing two-dimensional coherent states, we discover that GHA is capable of representing wave packets with Gaussian characteristics as they propagate along a trajectory closely resembling a classical periodic orbit over a brief period. We also show that it can be implemented in a non-separable billiard, the equilateral triangle, where we describe the algebra generators and the associated one-dimensional coherent states.

%Outline
The outline of this article is the following: In section \ref{sec: GHA}, we briefly review some basic concepts about the GHA together with this subspace description. Thus, we consider the eigenvalues obeying a second-order polynomial function. Then, we applied those concepts in section \ref{sec: application} for two different billiards: the square, in one and two dimensions, 
and an equilateral triangle, in one dimension. Finally, we state the conclusions in sec. \ref{sec: conclusion}.

%%%%%%%%%%%%%%%%%%%%%%%%%%%%%%%
\section{GHA}\label{sec: GHA}
\subsection{Brief description of GHA}
It is well known that the quantum harmonic oscillator can be described by the ladder operators $\hat{a}$ and $\hat{a}^\dagger$. The connection with the Hamiltonian is given by 
\begin{equation}\label{eq: harmonic oscilator}
\hat{H}_{osc} = \hbar \omega \left( \hat{a}^\dagger \hat{a} +\frac{1}{2}\right),     
\end{equation}
where $\omega$ is the natural frequency of the oscillator. Then, it leads to the definition of the number operator $\hat{n}=\hat{a}^\dagger \hat{a}$, and the usual commutation relations
\begin{eqnarray}\label{eq: comut QHSa}
    \left[\hat{a}, \hat{a}^\dagger\right]&=&1  \\ \label{eq: comut QHSb}
    \left[\hat{n}, \hat{a}^\dagger\right]&=&\hat{a}^\dagger  \\ \label{eq: comut QHSc}
    \left[\hat{n}, \hat{a}\right]&=&-\hat{a}  \ .
\end{eqnarray}
The operators $\hat{a}$, $\hat{a}^\dagger$ and $\hat{H}_{osc}$ together with the eq.\eqref{eq: comut QHSa}\eqref{eq: comut QHSb}\eqref{eq: comut QHSc} constitute an algebra, namely Heisenberg Algebra. In this context, the GHA was created with the objective of generalize this type of algebra to represents other quantum systems. The initial idea was to consider ladder operators $\hat{A}, \hat{A}^\dagger$ that connects to the Hamiltonian via
\begin{equation}\label{eq: Hdef}
    \hat{H}= \hat{A}^\dagger \hat{A}+c,
\end{equation}
where $c$ is a constant. Thus, the full development of the generalization was achieved by defining the algebra generators $\hat{H}, \hat{A}, \hat{A}^\dagger$ that satisfies

\begin{eqnarray}
\label{gen1}
 \hat{H} \hat{A}^\dagger & = & \hat{A}^\dagger f(\hat{H}) \\
 \label{gen2}
  \hat{A} \hat{H}  & = &  f(\hat{H}) \hat{A} \\
  \label{gen3}
  \left[ \hat{A}, \hat{A}^\dagger \right] &  = &  f(\hat{H}) - \hat{H}, \label{eq: comutador}
\end{eqnarray}
where $ \hat{A}=(\hat{A}^\dagger)^\dagger $ and the Hamiltonian is hermitian. This results in a characteristic function $f(.)$ that can represent the desired quantum system. The function $f(.)$ relates two consecutive energy eigenvalues ($\epsilon_{i+1}$, $\epsilon_i$) of the system  by means of the relation 
$\epsilon_{i+1} = f(\epsilon_i)$. One can observe that the function $f(\hat{H})=\hat{H}-1$ fulfill the comutation relations of the harmonic oscilator eq.\eqref{eq: comut QHSa}\eqref{eq: comut QHSb}\eqref{eq: comut QHSc}.

\subsection{GHA for the Hamiltonian's eigenstates subspace}\label{subsec: reviewgha}

We set the general vector $\vert l,m \rangle$ (which is an eigenstate of the Hamiltonian) to depict the Fock space n-dimensional representation theory. Therefore, $\hat{H} \vert l,m \rangle = \epsilon_{l,m} \vert l,m \rangle$, where $m$ and $l$ are quantum numbers, $\epsilon_{l,m} $ is the energy eigenvalue. It is worth mentioning that each state can be written as a tensor product of two Fock spaces 
 $\vert l, m \rangle = \vert l \rangle \otimes \vert m \rangle$.  
We construct an algebra for each quantum number $l$. Hence, for each algebra, we define the characteristic function as $f_l$, and $f_l^{(m)}(\epsilon_{l,1}) = \epsilon_{l,m}$, where $f_l^{(m)}$ is the m-th iterate of $\epsilon_{l,1}$. Hence, the operators $\hat{A}$ and $\hat{A}^\dagger$ are defined for each specific $l$, and its action on a general vector is
\begin{eqnarray} \label{eq: criacao}
 \hat{A}^\dagger \vert l,m \rangle & = & N_{l,m}\vert l,m+1 \rangle \\ \label{eq: aniquilacao}
 \hat{A} \vert l,m \rangle & = & N_{l,m-1}\vert l,m-1 \rangle,
\end{eqnarray}
where $N_{l,m}^2=\epsilon_{l,m+1}-\epsilon_{l,1}$. 

These generalized Heisenberg algebras describe several classes of quantum systems. It was presented in \cite{Curado2001b} that the class is characterized by quantum systems with the following relation
\begin{equation}
    \epsilon_{l,m+1} = f_l(\epsilon_{l,m}),
\end{equation}
where both $\epsilon$ are consecutive eigenvalues,  $f_l$ is the characteristic function in eq. (\ref{eq: comutador}), and $\hat{A}$ and $\hat{A}^\dagger$ are the annihilation and creation operators.

%%%%%%%%%%%%%%%%%%%%%%%%%%%%%%%
\subsection{Coherent states}
The GHA formulation provides an explicit form to obtain Klauder-type coherent states \cite{Hassouni2005} for each quantum number $l$,
\begin{equation}\label{eq: zbasico}
     \hat{A} \vert z,l \rangle = z \vert z,l \rangle,
 \end{equation}
 where $z$ is a complex number. Thus, one can expand $\vert z,l \rangle = \sum_{m=1}^\infty c_{l,m} \vert l,m \rangle$, and perform the action of the annihilation operator together with the eq. (\ref{eq: aniquilacao}) and (\ref{eq: zbasico}) 
 \begin{equation}
     \hat{A} \vert z,l \rangle = \sum_{m=1}^\infty c_{l,m+1} N_{l,m} \vert l,m \rangle = z \sum_{m=1}^\infty c_{l,m} \vert l,m \rangle,
 \end{equation}
 where  $\hat{A}  \vert l, 1 \rangle = 0$. Therefore, the coefficients $c_{l,m}$ are obtained via
 \begin{equation}
     c_{l,m} = c_{l,1} \frac{z^m}{N_{l,m-1}!} \, ,
 \end{equation}
 where $N_{l,m}!  \equiv N_{l,1}N_{l,2}\dots N_{l,m}$. Defining $N_l(z)=c_{l,1}$ and $N_{l,0}!=1$ for consistency, one can write the coherent states as 
 \begin{equation}\label{eq: coerenteprimeiro}
      \vert z,l \rangle = N_l(|z|)\sum_{m=1}^\infty \frac{z^m}{N_{l,m-1}!}\vert l,m \rangle.
 \end{equation}
 The Klauder's coherent states are obtained by the minimal set of conditions, which are the normalizability $\langle z \vert z \rangle=1$, continuity in the label 
 \begin{equation}
        |z-z'| \longrightarrow 0, \qquad \Vert \vert z \rangle - \vert z' \rangle \Vert \longrightarrow 0,
 \end{equation}
 and completeness
  \begin{equation}\label{eq: completeza}
     \vert l \rangle \langle l \vert \otimes \hat{\mathbb{I}}_m = \int \frac{d^2z}{\pi} \, \omega(z) \, \vert z,l \rangle \langle z,l \vert,
 \end{equation}
where $\omega(z)$ is a weight function, and in the left hand side of eq.\ref{eq: completeza} $\hat{\mathbb{I}}_m =\sum_m \vert m \rangle \langle m \vert$ is the identity operator. In this context, it is possible to construct coherent states belonging to a subset of all the eigenstates.  From the normalizability condition, one can obtain
 \begin{equation}\label{eq: normalizacao}
     N_l(|z|)= \left[\sum_{m=1}^\infty \frac{|z|^{2m}}{N^2_{l,m-1}!}\right]^{-1/2},
 \end{equation}
 and from the completeness condition, it is possible to find the weight function $\omega(z)$, for a given spectra.

\subsection{Second order polynomial} \label{subsubsec: secondorder}
In this section, we show the characteristic function of the algebra and construct the coherent states with one of the quantum numbers $l$ constant. We will consider the eigenvalues written as a second-order polynomial $\epsilon_{l,m}= a m^2+bm+c$, where $a$, $b$, and $c$ might depends of $l$ and let us assume $m = 1, 2, 3 \dots$ .
Thus, the function $f_l$ is obtained via the recurrence relation
\begin{equation}
     \epsilon_{l,m+1}= \epsilon_{l,m}+2 a m + a+b.
 \end{equation}
On the other hand, we can invert the second-order polynomial
\begin{equation}
     m = \frac{-b+\sqrt{b^2-4a(c-\epsilon_{l,m})}}{2a},
 \end{equation}
 where we take the positive sign due to the monotonically crescent behavior of the eigenvalues. Therefore,
 \begin{equation}\label{eq: funcCaracteristica}
    f_l(\epsilon_{l,m}) = \epsilon_{l,m+1}=a+\epsilon_{l,m} +\sqrt{b^2-4a(c-\epsilon_{l,m})},
 \end{equation}
 where $f_l$ is the characteristic function.
It is important to note that any characteristic function has the same form as eq.(\ref{eq: funcCaracteristica}) if the eigenvalues obey a second-order polynomial function. 

 The normalization factor $N_l(z)$ in eq. (\ref{eq: normalizacao})  can be obtained with the help of the second order polynomial, where $N_{l,m}^2=\epsilon_{l,m+1}-\epsilon_{l,1} = am^2+(2a+b)m$,
 \begin{equation}
     N_l(|z|)= \left[\sum_{m=1}^\infty \frac{|z|^{2m}}{\prod_{n=1}^{m-1}[an^2+(2a+b)n]}\right]^{-1/2}.
 \end{equation}
 The denominator $N_{l,m-1}^2!$ can be simplified
 \begin{equation}
 \label{eq: NFatorial}
  N_{l,m-1}^2! =  \prod_{n=1}^{m-1}[an^2+(2a+b)n]  = \frac{a^{m-1}(m-1)! \, \Gamma\left(2+\frac{b}{a} +m\right) }{\Gamma\left(3+\frac{b}{a}\right)} \,\,, 
 \end{equation}   
 therefore,
  \begin{equation}
     N_l(|z|)= \left[\sum_{m=1}^\infty \frac{|z|^{2m} \Gamma\left(3+\frac{b}{a}\right) }{a^{m-1}(m-1)! \, \Gamma\left(2+\frac{b}{a} +m \right) }\right]^{-1/2},
 \end{equation}
 where $\Gamma$ is the Gamma function. Setting $k=m-1$
  \begin{equation}
     N_l(|z|)= \left[\frac{\Gamma\left(3+\frac{b}{a}\right) |z|^{-\frac{b}{a}}}{(\sqrt{a})^{-2-\frac{b}{a}}}\sum_{k=0}^\infty \frac{|z|^{2k+2+\frac{b}{a}}}{k! \, (\sqrt{a})^{2k+2+\frac{b}{a}} \ \Gamma\left(3+\frac{b}{a} +m\right) }\right]^{-1/2},
 \end{equation}
 we recognize the sum as the modified Bessel function $I_\nu$ (eq. 8.445 \cite{gradshteyn2014table})
  \begin{equation}
  \label{eq: normaliFIM}
     N_l(|z|) = \left[ a \left( \frac{\sqrt{a}}{|z|}\right)^{\frac{b}{a}} \Gamma\left( 3+\frac{b}{a}\right) I_{2+b/a}(2|z|/\sqrt{a})\right]^{-1/2}.
 \end{equation}
Along these lines, the weight function $\omega (z)$ is obtained via the integration in eq. (\ref{eq: completeza}) over the complex plane setting $z=\rho \ \exp{i \theta}$, 

\begin{equation}
 \int \frac{d^2z}{\pi}  \, \omega(z) \vert z,l \rangle \langle z,l \vert =\sum_{m,n=1}^\infty \frac{\vert l,m \rangle \langle l,n \vert }{ 2\pi N_{l,m-1}!N_{l,n-1}! } \int_0^\infty \int_0^{2\pi} \rho^{m+n} e^{i \theta(m-n)}\, \omega(\rho)\,N_l^2(\rho)\, \rho \, d\theta d\rho ,
\end{equation}
 thus, with the help of eq. (\ref{eq: normaliFIM}) and the orthogonality of the complex exponential function
 \begin{equation}
    \int \frac{d^2z}{\pi}  \, \omega(z) \vert z,l \rangle \langle z,l \vert =\sum_{m=1}^\infty \frac{\vert l,m \rangle \langle l ,m \vert}{ N_{l,m-1}^2! } \int_0^\infty  \frac{\rho^{2m+1} \, \omega(\rho)}{a \left( \frac{\sqrt{a}}{\rho}\right)^{\frac{b}{a}} \Gamma\left( 3+\frac{b}{a}\right) I_{2+b/a}(2\rho/\sqrt{a})} d\rho ,
\end{equation}
and substituting $N_{l,m-1}^2!$, defining $x=\rho/\sqrt{a}$, and factoring out the common terms,  
 \begin{equation}\label{eq: pesorelacao}
    \int \frac{d^2z}{\pi}  \, \omega(z) \vert z,l \rangle \langle z,l \vert = a \, \sum_{m=1}^\infty \frac{ \vert l,m \rangle \langle l,m \vert}{ (m-1)! \,\Gamma\left(m+2+\frac{b}{a}\right) } \int_0^\infty  \frac{x^{2m+1+\frac{b}{a}} \, \omega(x)}{ I_{2+b/a}(2x)} dx .
\end{equation}.  
One can verify the weight function
\begin{equation}
    \omega(x) = \frac{4}{a}I_{2+b/a}(2x)K_{2+b/a}(2x),
\end{equation}
solves the eq. (\ref{eq: pesorelacao}) with eq. 6.561.16 of \cite{gradshteyn2014table}. Therefore,  
\begin{equation}
\label{eq: funcaopeso1d}
    \omega(|z|) = \frac{4}{a} I_{2+b/a}\left( \frac{2|z|}{\sqrt{a}} \right) \, K_{2+b/a}\left(\frac{2|z|}{\sqrt{a}}\right).
\end{equation} 

\noindent
These analytical results will be used in the next examples of billiards.

%%%%%%%%%%%%%%%%%%%%%%%%%%%%%%%%
\section{Quantum Billiards}\label{sec: application}
Quantum billiards refer to the study of quantum mechanical systems where a particle is confined within a bounded region. We select two representative cases: the square billiard (subsec. \ref{subsec: quadrado}) and the equilateral triangle (subsec.\ref{subsec: equilatero}). The square billiard is defined as separable billiards. The meaning behind this classification is that their eigenstates can be found via the separation of variables method. On the other hand, the equilateral triangle is an example of a nonseparable billiard. Also, a second-order polynomial gives the eigenvalues of the square and the equilateral billiard. Along these lines, we determine the algebra generators and their position representation, we construct coherent states and verify their quantum time revival.

\subsection{Square Billiard}\label{subsec: quadrado}
The dynamics of a particle of mass $\mu$, 
confined in a rectangular billiard,  are governed by the Schrodinger equation with the potential
\begin{equation}
     V(\textbf{r)}= 
    \begin{cases} 
      0 & 0< x < L_x  \ {\rm and} \ 0< y <L_y \\
      \infty & {\rm otherwise}
   \end{cases}
   ,
 \end{equation}
 therefore the particle can not be found outside of the rectangle of sides $L_x$ and $L_y$. Following the separation of variables method, one can find the energy eigenvalues
\begin{equation}
\label{eigensquare}
    \epsilon_{l,m}=\frac{\hbar^2\pi^2}{2\mu}\left(\frac{l^2}{L_x^2}+\frac{m^2}{L_y^2}\right),
\end{equation}
where $l,m=1,2,3...$ associated with the eigenfunctions
\begin{equation}
    \psi_{l,m}(\textbf{r})=\frac{2}{\sqrt{L_xL_y}}\sin\left(\frac{l\pi x}{L_x}\right)\sin\left(\frac{m\pi y}{L_y}\right).
\end{equation}
Setting $\hbar=2\mu = L_x=L_y=1$, we depict the square billiard. In this context, we associate the eigenfunction $\psi_{l,m}$ in the Hilbert space with each state $\vert l,m \rangle$ in Fock state space. Then, one can easily see that the characteristic function  (accordingly to eqs. \ref{eq: funcCaracteristica}) and \ref{eigensquare} is
\begin{equation}
    f_l(\epsilon_{l,m}) =\pi^2+\epsilon_{l,m} +\sqrt{4\pi^2\epsilon_{l,m}-4l^2\pi^4} \,\, .
\end{equation}
Hence, the actions of the algebra  generators, see eqs. \ref{gen1}, \ref{gen2}, \ref{gen3} , are
\begin{eqnarray}
    \hat{H}_m \vert l,m \rangle & = & \pi^2 m^2\vert l,m \rangle, \\
    \hat{A}^\dagger \vert l,m \rangle & = & \pi\sqrt{m(m+2)} \vert l,m+1 \rangle, \\
    \hat{A} \vert l,m \rangle & = & \pi \sqrt{m^2-1} \vert l,m-1 \rangle,
\end{eqnarray}
 where $\hat{H}_m$ represents a separable part of the Hamiltonian, which is related to the state's subspace $\vert m \rangle$. In other words, the Hamiltonian of the system is $\hat{H}=\hat{H}_l \otimes \hat{H}_m$. It is easy to check that $\hat{H}_m = \hat{A}^\dagger \hat{A}+\pi^2$, which agrees with eq. \eqref{eq: Hdef}.
 
 The ladder operators in the position representation given by, 
\begin{equation}\label{eq: estruturaadaga}
    \hat{A}^\dagger \psi_{l,m}(\textbf{r})= \left[g_1(y)\frac{d}{dy}+g_2(y)\rho(\hat{N})\right]\tau(\hat{N}) \, \psi_{l,m}(\textbf{r})=N_{l,m} \,\psi_{l,m+1}(\textbf{r}),
\end{equation}
where
\begin{eqnarray}
     g_1(y)&=&\sin(\pi y) \\
     g_2(y) &=& \cos(\pi y) \\
     \rho(\hat{N})& = & \hat{N} \pi \\
     \tau(\hat{N}) &=& \frac{\sqrt{\hat{N}(\hat{N}+2)}}{\hat{N}},
 \end{eqnarray}
 and
 \begin{equation}
     \hat{A}=\tau(\hat{N})\left[-\frac{d}{dy}g_1(y)+\rho(\hat{N})g_2(y)\right],
 \end{equation}
where $\hat{N}$ is the number operator, $\hat{N} \vert l, m \rangle = m \vert l, m \rangle$. So, it yields the commutation relations, 
\begin{eqnarray}
    \left[ \hat{H}_m, \hat{A}^\dagger \right] &  = &  \hat{A}^\dagger \left( \pi^2 \, \hat{\mathbb{I}} + 2\pi^2 \hat{N}\right) \\
    \left[ \hat{H}_m, \hat{A} \right] &  = &  -\left( \pi^2\hat{\mathbb{I}}+2\pi^2\hat{N}\right)\hat{A} \\
    \left[ \hat{A}, \hat{A}^\dagger\right] &  = &  \pi^2\hat{\mathbb{I}}+2\pi^2\hat{N} \,\, .
\end{eqnarray} 

%%%%%%%%%%%%%%%%%%%%%%%%%%%%%%%%%%%%
\subsubsection{Coherent States}
\begin{figure}[htp!]
    \centering
    \includegraphics[width=.5\textwidth]{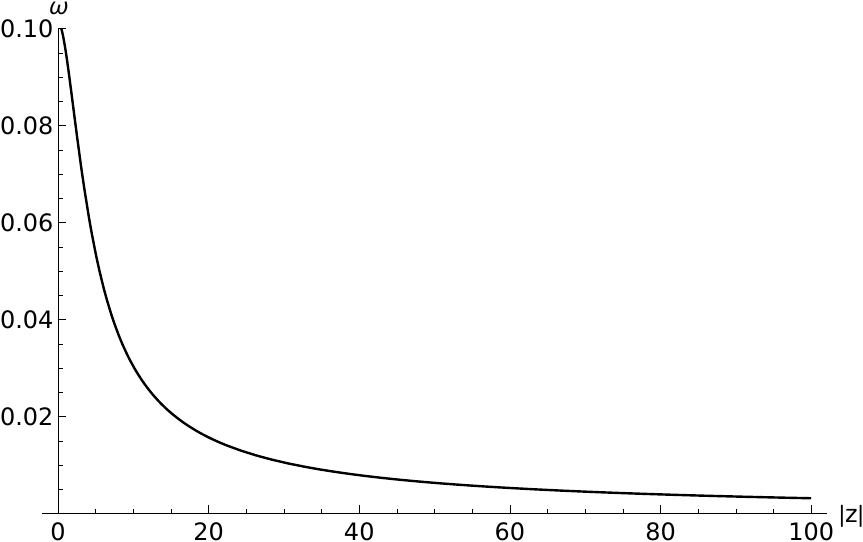}
    \caption{Plot of the weight function (eq. \ref{eq: funcaopeso1dquadrado}) related to the} one-dimensional coherent state of the square billiard.
    \label{Fig: pesoquadrado1d}
\end{figure}

The coherent states related to the characteristic function in the form of eq. \ref{eq: funcCaracteristica} is denominated as nonlinear coherent states \cite{Hassouni2005}. In this work, we only refer to it as coherent states. Then, we define the ``one dimensional'' coherent state $( \ \vert z,l \rangle \ )$ as the cases where one of the quantum numbers $(l)$ is fixed. Therefore, they are constructed by the GHA, which depicts a subspace of all eigenfunctions. The denominator becomes, see eq. \ref{eq: NFatorial}, with $a=\pi^2, b=c=0$, 
\begin{equation}
    N_{l,m-1}^2! = \frac{(m-1)!(m+1)!(\pi^2)^{m-1}}{2},
\end{equation}
and the normalization, see eq. \ref{eq: normaliFIM} with $a=\pi^2, b=c=0$, is 
\begin{equation}
    N_{l}(|z|) = \left[ 2\pi^2 I_2(2|z|/\pi ) \right]^{-1/2},
\end{equation}
which gives the one-dimensional coherent state
\begin{equation}\label{eq: EstCoeQuadUmDim}
    \vert z,l \rangle = \frac{1}{\sqrt{I_2(2|z|/\pi )}}\sum_{m=1}^\infty \left(\frac{z}{\pi}\right)^m \frac{1}{\sqrt{(m-1)! \, (m+1)!}}\vert l,m \rangle.
\end{equation}
It is worth mentioning that the corresponding weight 
function, see eq. \ref{eq: funcaopeso1d}, shown in (Fig. \ref{Fig: pesoquadrado1d}), is given by,  
\begin{equation}\label{eq: funcaopeso1dquadrado}
    \omega(|z|) = \frac{4}{\pi^2} I_{2}\left(\frac{2|z|}{\pi}\right)K_{2}\left(\frac{2|z|}{\pi}\right).
\end{equation}. 

The time evolution of the coherent state $\vert z,l,t \rangle = e^{-i \hat{H}t} \vert z,l \rangle$ can be trivially found
\begin{equation}
    \vert z,l,t \rangle = \frac{1}{\sqrt{I_2(2|z|/\pi )}}\sum_{m=1}^\infty \left(\frac{z}{\pi}\right)^m \frac{e^{-i t \pi^2 (l^2+m^2)}}{\sqrt{(m-1)! \,(m+1)!}}\vert l,m \rangle.
\end{equation}
This allows us to 
determine the fidelity $\mathcal{F}_l(z,t)$ between the initial state $\vert z,l\rangle$ and the time evolved one $\vert z,l,t \rangle$,
\begin{equation}
    \mathcal{F}_l(z,t)=\vert \langle z,l\vert z,l,t \rangle \vert^2= \frac{1}{I_2(2|z|/\pi) } \, \left| \sum_{m=1}^\infty \left(\frac{|z|}{\pi}\right)^{2m}   \frac{e^{-i t \pi^2 m^2}}{(m-1)! \, (m+1)! } \right|^2.
\end{equation} 

In position representation ($z = \rho e^{i \theta}$), we have the coherent state
\begin{equation}
\label{eq: CSquadrado1}
 \mathcal{Z}_l(\textbf{r}, \rho,\theta,t) 
=\frac{1}{\sqrt{I_2(2 \, \rho /\pi )}}\sum_{m=1}^\infty \left(\frac{\rho e^{i \theta}}{\pi}\right)^m \frac{e^{-i t \pi^2 (l^2+m^2)}}{\sqrt{(m-1)! \, (m+1)!}}\psi_{l,m}(\textbf{r}).
\end{equation}  
\begin{comment}
As an illustrative example, we set $z=5$ and $l=1$ for the coherent state $\vert 5,1,t \rangle$ and perform a plot (Fig. \ref{Fig: fidquad}a) of the fidelity over a quantum revival time \cite{Robinett2004} $T_{rev} = 4\mu L^2/ \hbar \pi = 2/\pi \approx 0.64$.
In position representation, we have the coherent state
\begin{equation}\label{eq: CSquadrado1}
\mathcal{Z}_l(\textbf{r},z,t)=\frac{1}{\sqrt{I_2(2|z|/\pi )}}\sum_{m=1}^\infty \left(\frac{z}{\pi}\right)^m \frac{e^{-i t \pi^2 (l^2+m^2)}}{\sqrt{(m-1)!(m+1)!}}\psi_{l,m}(\textbf{r}),
\end{equation}
then we show density plots (Fig. \ref{Fig: fidquad}b-j) of the probability density of the coherent state $\mathcal{Z}_1(\textbf{r},5,t)$ for 8 time steps in the range $t=0$ and $t=T_{rev}/2$. Through all this work we truncate the coherent states at the 20-th term.

\end{comment}

%%%%%%%%%%%%%%%%%%%%%%%%%%%%%%%%%%%%%%%%

\subsubsection{Two-dimensional coherent state}

\begin{figure}
    \centering
    \includegraphics[width=.6\textwidth]{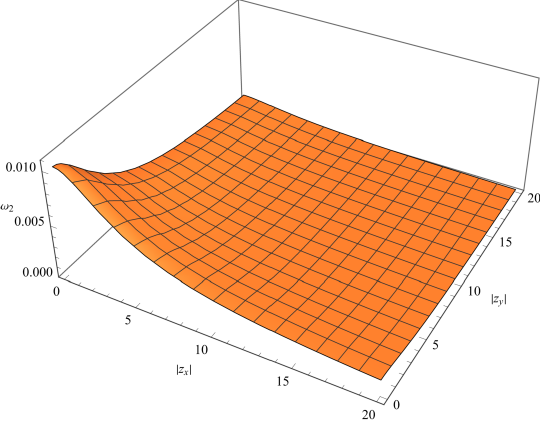}
    \caption{3D plot of the weight function $\omega_2(\textbf{z})$ (eq. \ref{eq: peso2}) related to the two-dimensional coherent state of the square billiard.}
    \label{Fig: pesoquadrado2d}
\end{figure}

\noindent
The one-dimensional coherent state (eq. \ref{eq: EstCoeQuadUmDim}) was constructed considering only a subspace of the eigenstates, leaving one of the quantum numbers constant. Therefore, it can be written as, 
\begin{equation}
     \vert z,l \rangle = \vert l \rangle \otimes 
 \left[ \frac{1}{\sqrt{I_2(2|z|/\pi )}}\sum_{m=1}^\infty \left(\frac{z}{\pi}\right)^m \frac{1}{\sqrt{(m-1)! \, (m+1)!}}\vert m \rangle \right],
\end{equation}
because the normalization constant and the denominator does not depend on the quantum number $l$. Then, we write the two-dimensional coherent state as a tensor product 
\begin{eqnarray} \label{eq: EstCoeQuadDoisDim}
    \vert \textbf{z} \rangle &=& \left[ \frac{1}{\sqrt{I_2(2|z_x|/\pi )}}\sum_{l=1}^\infty \left(\frac{z_x}{\pi}\right)^l \frac{1}{\sqrt{(l-1)! \, (l+1)!}}\vert l \rangle \right] \otimes \\ & &  
 \left[ \frac{1}{\sqrt{I_2(2|z_y|/\pi )}}\sum_{m=1}^\infty \left(\frac{z_y}{\pi}\right)^m \frac{1}{\sqrt{(m-1)! \, (m+1)!}}\vert m \rangle \right],
\end{eqnarray}
 with $\textbf{z}=(z_x,z_y)$,  where both, $z_x$ and $z_y$, are complex numbers. This state is normalized and it obeys the completeness relation
\begin{equation}
\mathbb{I}=\mathbb{I}_l\otimes\mathbb{I}_m =\int \frac{d^2z_x}{\pi} \frac{d^2z_y}{\pi}  \ \omega_2(\textbf{z}) \vert \textbf{z} \rangle \langle \textbf{z} \vert, 
\end{equation} 
where the weight function 
\begin{equation}\label{eq: peso2}
    \omega_2(\textbf{z})=\omega(|z_x|) \,\omega(|z_y|),
\end{equation}
$\omega(|z|)$ was presented at eq.\ref{eq: funcaopeso1d}, and the corresponding plot is presented at Fig. \ref{Fig: pesoquadrado2d}. 

The eigenfunctions $\psi_{l,m}(\textbf{r},t)$ are separable ($\textbf{r}=(x, y)$), in such a way that one can interpret them as the product of two eigenfunctions $\phi_l(x,t)=\sqrt{2}\sin(l \pi x)\exp{(-i\pi^2l^2 t)}$  of a particle in an infinite quantum well of unitary width, 
\begin{equation}\label{eq: psi2d}
    \psi_{l,m}(\textbf{r},t)=\phi_l(x,t) \, \phi_m(y,t) \, .
\end{equation}
Therefore, the coherent state $\mathcal{Z}_l(\textbf{r},z_y,t)$ (eq. \ref{eq: CSquadrado1}, where $\rho$ and $\theta$ is depicted as $z_y$, and eq. \ref{eq: psi2d}) in position representation is also separable
\begin{equation}\label{eq: CSquadrado2}
    \mathcal{Z}_l(\textbf{r},z_y,t)=\phi_l(x,t)\left[ \frac{1}{\sqrt{I_2(2|z_y|/\pi )}}\sum_{m=1}^\infty \left(\frac{z_y}{\pi}\right)^m \frac{1}{\sqrt{(m-1)!(m+1)!}}\phi_{m}(y,t)\right] \, .
\end{equation}
 Hence, we can write the two-dimensional coherent states in terms of eigenfunctions 
\begin{eqnarray}\label{eq: coherentstate2dPosition}
\mathcal{Z}_{2D}(\textbf{r},\textbf{z},t) & =& \left[ \frac{1}{\sqrt{I_2(2|z_x|/\pi )}}\sum_{l=1}^\infty \left(\frac{z_x}{\pi}\right)^l \frac{1}{\sqrt{(l-1)!(l+1)!}}\phi_{l}(x,t)\right] \times  \\ 
    & & \left[ \frac{1}{\sqrt{I_2(2|z_y|/\pi )}}\sum_{m=1}^\infty \left(\frac{z_y}{\pi}\right)^m \frac{1}{\sqrt{(m-1)!(m+1)!}}\phi_{m}(y,t)\right], \nonumber
\end{eqnarray}
where $\textbf{z}=(z_x,z_y)$. The fidelity can be readily obtained
\begin{eqnarray}
    \mathcal{F}_{2D}(t) & = & \left|\int_0^1\int_0^1 \mathcal{Z}_{2D}^*(\textbf{r},\textbf{z},0)\mathcal{Z}_{2D}(\textbf{r},\textbf{z},t)dx dy \right|^2 \\ 
    & = & \left|\frac{1}{I_2(2|z_x|/\pi )I_2(2|z_y|/\pi )}\sum_{l,m=1}^\infty \frac{|z_x|^{2l}|z_y|^{2m}}{\pi^{2l+2m}} \frac{e^{-i t \pi^2 (l^2+m^2)}}{(m+1)!(m-1)!(l+1)!(l-1)!} \right|^2. \nonumber
\end{eqnarray}

\begin{figure}[htp!]
    \centering
    \includegraphics[width=.8\textwidth]{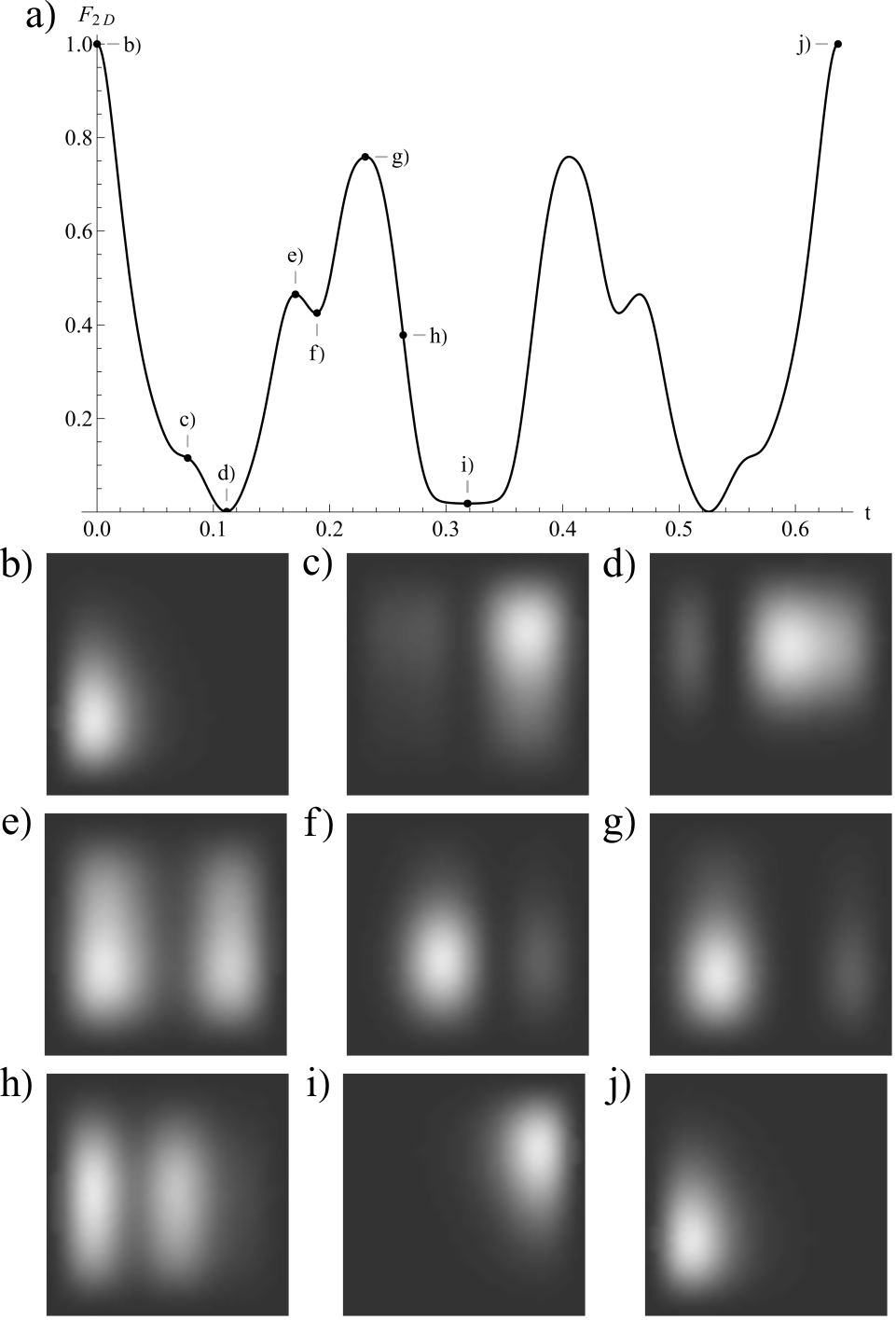}
    \caption{Plot of the fidelity (a) between the initial coherent state $\mathcal{Z}_{2D}(\textbf{r},\textbf{z},0)$ and the evolving one $\mathcal{Z}_{2D}(\textbf{r},\textbf{z},t)$ over a quantum revival time $T_{rev}$. The dots indicate the time for each specific density plot (b-j) of the probability density of the time-evolved two-dimensional coherent state in position representation $|\mathcal{Z}(\textbf{r},z,t)|^2$ (eq \ref{eq: coherentstate2dPosition}), where $z_x=5$, $z_y=2$, and we truncate each series with 20 terms. Note that at the point $j$ the fidelity reaches one.}
    \label{Fig: fidquadcoherent2D}
\end{figure}

As an illustrative example, we set $\textbf{z}=(5,2)$ for the coherent state eq. \ref{eq: coherentstate2dPosition} and perform a plot (Fig. \ref{Fig: fidquadcoherent2D}a) of the fidelity over a quantum revival time \cite{Robinett2004} $T_{rev} = 4\mu L^2/ \hbar \pi = 2/\pi \approx 0.64$, where $L=L_x=L_y=1$.

%%%%%%%%%%%%%%%%%%%%%%%%%%%%%%%
\subsubsection{A classical correspondence}

\begin{figure}
    \centering
    \includegraphics[width=.8\textwidth]{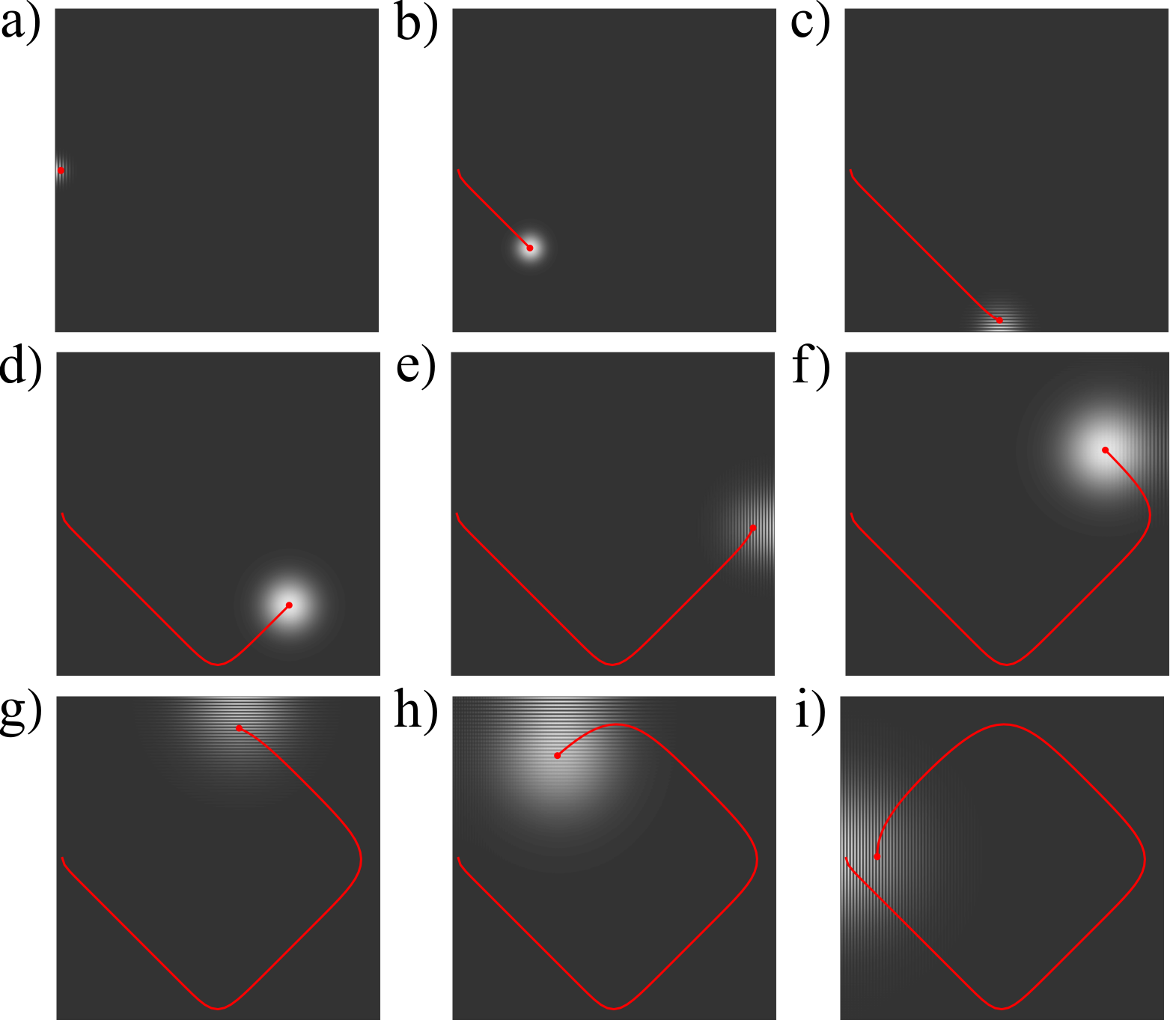}
    \caption{Density plot (a-i) of the probability density of the two-dimensional coherent state (eq. \ref{eq: ECOETEMP}) at times (0, 3, 6, 9, 12, 15, 18, 21, 25) $T_{rev}/5000$ respectively. The red dot indicates the point $\mathbf{r}$ related to the expected value position, while the red curves the past values since $t=0$. }
    \label{Fig: evotempposesp}
\end{figure}

\begin{figure}
    \centering
    \includegraphics[width=.49\columnwidth]{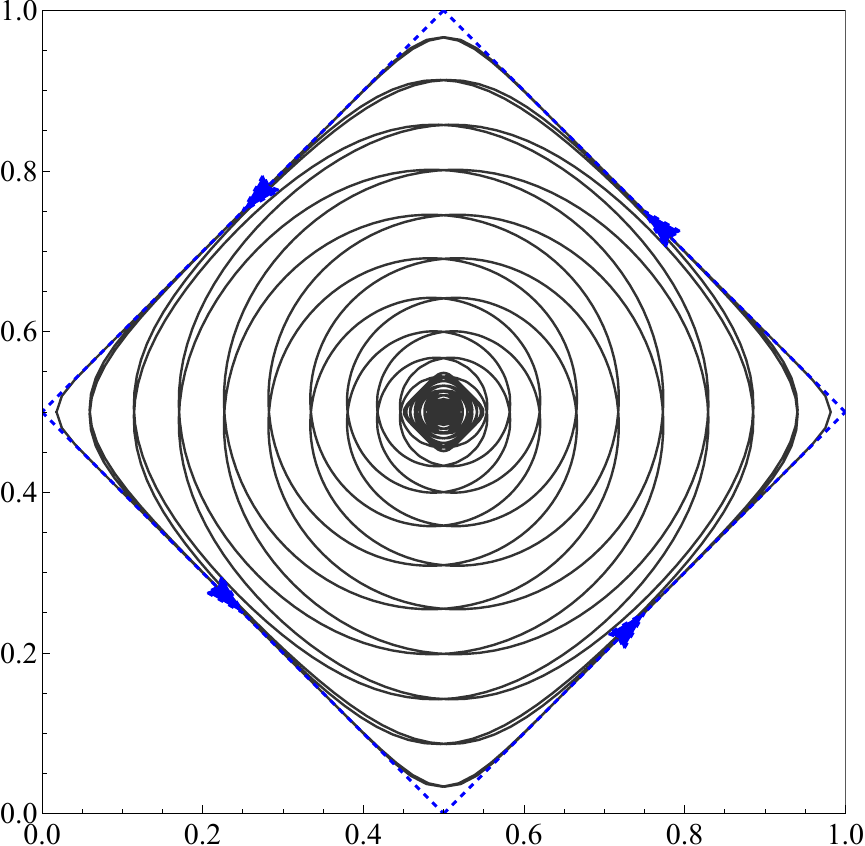}
    \caption{(Black) Trajectory of $\mathbf{r}$ for the coherent state (eq. \ref{eq: ECOETEMP}) with $\mathbf{z}=(100\pi,100\pi i)$ for a full quantum revival time $T_{rev}$ cycle.(Dashed blue arrows) A classical trajectory. For a brief time, the expected value of the position behaves close to the classical periodic orbit.}
    \label{fig: fulltraj}
\end{figure}

The quantum states associated with classical periodic orbits are superpositions of nearly degenerate eigenstates. Along these lines, coherent states were designed to depict the classical periodic orbits \cite{sym15101809}, where the classical equations of motion can be extracted from the wave functions. However, the form of wave function drastically changes with time, and visually resembles the classical trajectory at a specific time. Here, using the two-dimensional coherent states we find that GHA can depict Gaussian-like wave packets traveling in a path close to a classical periodic orbit for a small time interval. One can observe in fig. \ref{Fig: evotempposesp} the trajectory of the expected value of the position operator. The red dot corresponds to the point
\begin{equation}
   \langle \mathbf{\hat{r}} \rangle_z = \left( \langle \hat{x}\rangle_{\mathbf{z}}, \langle \hat{y} \rangle_{\mathbf{z}}\right),
\end{equation} 
where $\langle\hat{\bullet}\rangle_{\mathbf{z}}$ is the expected value of a desired operator calculated for the two-dimensional coherent state 
\begin{equation} \label{eq: ECOETEMP}
    \vert \mathbf{z},t \rangle  =\frac{1}{\left[ I_2(2|z_x|/\pi )I_2(2|z_y|/\pi )\right]^{\frac{1}{2}}}\sum_{l,m=1}^\infty \left(\frac{z_x}{\pi}\right)^l\left(\frac{z_y}{\pi}\right)^m \frac{e^{-i t \pi^2 (l^2+m^2)}}{\left[ (l-1)! \, (l+1)! \, (m-1)! \, (m+1)!\right]^{\frac{1}{2}} }\vert l,m \rangle,
\end{equation}
with $\mathbf{z}=(100\pi,100\pi i)$, at a given time $t$. It is easy to check that the expected value of $\hat{x}$ is, 
\begin{equation}
    \langle \hat{x}\rangle_{\mathbf{z}} =\frac{1}{ I_2(2|z_x|/\pi )}\sum_{l,l'=1}^\infty\left(\frac{z_x}{\pi}\right)^l\left(\frac{z_x^*}{\pi}\right)^{l'}\frac{c_{l,l'}}{\left[ (l-1)! \, (l+1)! \, (l'-1)! \, (l'+1)!\right]^{\frac{1}{2}}},
\end{equation}
where 
\begin{equation}
    c_{l,l'}= \begin{cases} 
      \frac{L}{2} & l= l' \\
      0 & l+l' \ {\rm even} \\
      \frac{-8L l l'}{\pi^2 (l^2-l'^2)^2} & l+l' \ {\rm odd} 
   \end{cases}  \, .
\end{equation}
A similar expression to $\langle \hat{y}\rangle_{\mathbf{z}}$ can be obtained with the replacement $z_x \rightarrow z_y$. Therefore, in fig \ref{Fig: evotempposesp} we show a density plot of the probability density of the coherent state (eq. \ref{eq: ECOETEMP}) for different times in the interval $t \in \left[0, T_{rev}/200 \right]$, and the red line corresponds to the past values of $\textbf{r}$ since the beginning. It is important to note that the coefficients that contribute the most to the sum in eq. \ref{eq: ECOETEMP} are the ones around $\Bar{l} = \vert z_x \vert \pi$ and $\Bar{m} = \vert z_y \vert \pi$, in this case $\Bar{l}=\Bar{m}=100$. Consequently, we choose to truncate the sum within $70\leq l,m \leq 130$. We examined a broader range of indices $l,m$ and observed that the wave function remains unchanged. The Gaussian-like wave packet spreads with time, but at the time around $t= T_{rev}/2$, the wave packet reconstructs its initial form in the right side of the billiard, then starts to spread again. The wave packet restores itself one more time at time $t=T_{rev}$ achieving the unit fidelity with the initial state. In fig. \ref{fig: fulltraj} we plot the trajectory of $\langle \mathbf{\hat{r}} \rangle_z $ for a $T_{rev}$ cycle, and it is possible to observe that for a short period, the trajectory is close to the classical periodic orbit. We state that one can construct a similar pattern for other periodic orbits by adjusting the values of $\mathbf{z}$

%%%%%%%%%%%%%%%%%%%%%%%%%%%%%%%%%%%%
\subsection{Equilateral Triangle Billiard}\label{subsec: equilatero}
\begin{figure}[htp!]
     \centering
     \includegraphics[width=.45\textwidth]{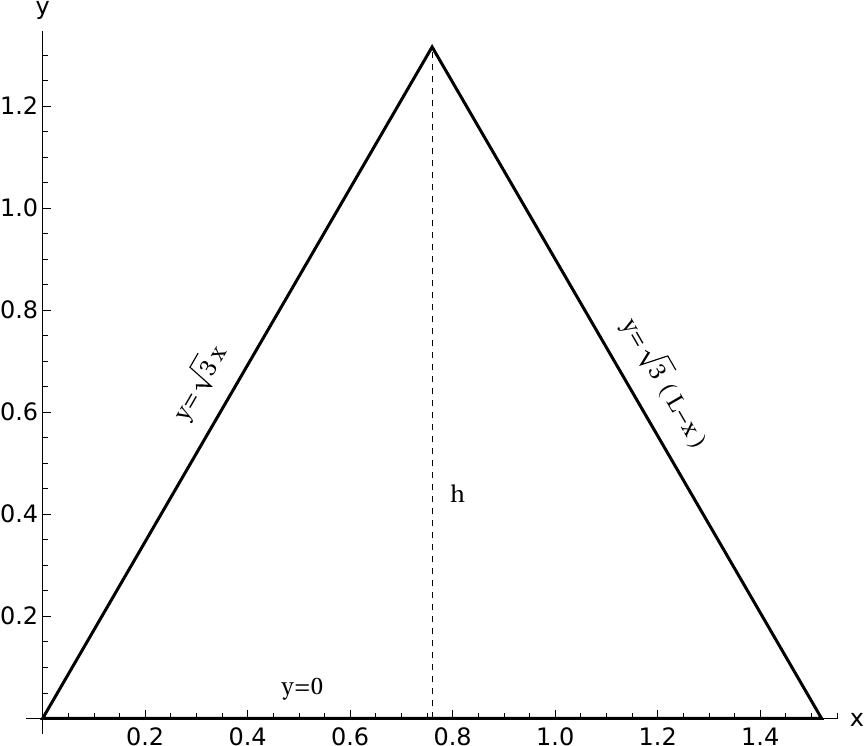}
     \includegraphics[width=.45\textwidth]{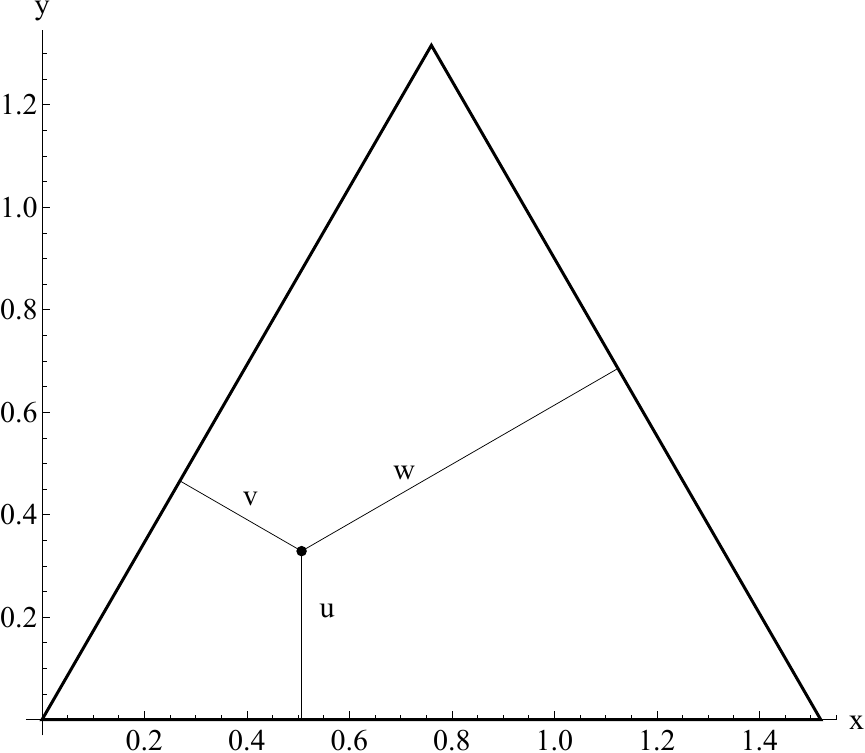}
     \caption{(Left) Equilateral triangle billiard with unit area, side's length $L \approx 1.52$, and height $h\approx 1.32$. (Right) A schematic portrayal of the auxiliary variables $u$, $v$, and $w$ that are proportional to the perpendicular distances of the triangle's sides.}
     \label{fig: trianguloequilatero}
 \end{figure}

The Hamiltonian of the system is $\hat{H}= -\nabla^2 +V(\textbf{r})$, where we considered $\hbar=2\mu=1$, and the potential that models the equilateral triangle billiard
\begin{equation}
     V(\textbf{r})= 
    \begin{cases} 
      0 & \textbf{r} \in \Omega \\
      \infty & {\rm elsewhere} \, \, ,
   \end{cases}
 \end{equation}
 where $\Omega$ is an interior region defined by the intersection of $y \leq \sqrt{3}x$, $y\leq \sqrt{3}(L-x)$ and $y\geq 0$ (Fig. \ref{fig: trianguloequilatero}). We set the area of the triangle to be unit, then the side's length is $L=\approx 1.52$, and the height $h \approx 1.32$
 
 The analytical solution of the wave function can not be obtained by the separation of variables. However, it can be found by introducing the auxiliary variables 
 \begin{eqnarray}
     u &=& \left(\frac{2\pi}{h}\right)y\\
     v &=& \left(\frac{2\pi}{h}\right)\left(-\frac{y}{2}+\frac{\sqrt{3}x}{2}\right) \\
     w &=& \left(\frac{2\pi}{h}\right)\left(-\frac{y}{2}-\frac{\sqrt{3}x}{2}\right) +2\pi,
 \end{eqnarray}
where $h$ is the triangle altitude. As indicated in Fig. \ref{fig: trianguloequilatero}, they are proportional to the perpendicular distance to the edges, obey the relation $u+v+w=2\pi$, and allow us to write the boundary conditions in a symmetric form,
\begin{equation}
    \psi(\textbf{r})=0 , \ {\rm when} \ 
    \begin{cases} 
      u=0 & v=2\pi - w \\
      v=0 & w=2\pi - u \\
      w=0 & u=2\pi - v 
   \end{cases}
   .
\end{equation}
    The point group $C_{3v}$ maintains the invariance of the equilateral triangle problem, and this group is isomorphic to the symmetric group $S_3$. Along these lines, these variables transform under the $S_3$ group, which can be understood as permutation operations $(\sigma_1,\sigma_2, \sigma_3)$ and rotation operations $(I,C_3,C_3^2)$ of the vertices/sides, where $I$ is the identity. To represent the variable transformations we can define the vector $(u,v,w)^T$. Then, it transforms with the following matrices,
 \begin{equation}
     I = 
    \begin{pmatrix}
        1 & 0 & 0\\
        0 & 1 & 0 \\
        0 & 0 & 1
    \end{pmatrix}
, \qquad C_3 = 
    \begin{pmatrix}
        0 & 1 & 0\\
        0 & 0 & 1 \\
        1 & 0 & 0
    \end{pmatrix}
, \qquad C_3^2 =     
    \begin{pmatrix}
        0 & 0 & 1\\
        1 & 0 & 0\\
        0 & 1 & 0
    \end{pmatrix}
,
 \end{equation}
 \begin{equation}
     \sigma_1 = 
    \begin{pmatrix}
        0 & 1 & 0\\
        1 & 0 & 0 \\
        0 & 0 & 1
    \end{pmatrix}
, \qquad \sigma_2 = 
    \begin{pmatrix}
        0 & 0 & 1\\
        0 & 1 & 0 \\
        1 & 0 & 0
    \end{pmatrix}
, \qquad \sigma_3 =     
    \begin{pmatrix}
        1 & 0 & 0\\
        0 & 0 & 1\\
        0 & 1 & 0
    \end{pmatrix}
.
 \end{equation}
One can observe that the number of transformations is higher than the order of the matrices. Therefore, it implies the existence of three irreducible representations for this group, which are denominated as $A_1$, $A_2$, and $E$ (for further details see \cite{Li1985, Li1987, Krishnamurthy1982, Turner1984}). In this work, we will consider only the irreducible representation $A_1$, because it will be related to one family of eigenfunctions, in other words, with the same quantum number $q=0$. The usual procedure to obtain the analytic solutions for the wave function is applying a specific projector operator in a basis function $f(p u -q v)$, where $p$ and $q$ are constants, and $f$ can be the sine, cosine, or exponential functions. Henceforth, the specific operator is the one related to $A_1$ representation, which is the projector defined as $\mathcal{P}(A_1)=I+C_3+C_3^2 + \sigma_1+\sigma_2+\sigma_3$. Thus, when it is applied to the basis function it generates
\begin{equation}
    \psi_{p,q}(\textbf{r})= f(pu-qv) +f(pv-qw)+f(pw-qu)+f(pv-qu)+f(pw-qv)+f(pu-qw), 
\end{equation}
choosing $f$ as the sine function we have
\begin{eqnarray}
    \psi_{p,q}(\textbf{r}) &=& \cos \left[q \sqrt{3}\pi x/h\right] \sin \left[(2p+q) \pi y/h\right] \nonumber \\
    & & -\cos \left[p \sqrt{3}\pi x/h\right] \sin \left[(2q+p) \pi y/h\right] \nonumber \\
    & & -\cos \left[ (p+q) \sqrt{3}\pi x/h\right] \sin \left[(p-q) \pi y/h\right],
\end{eqnarray}
    where $q=0,1,2...$ and $p=q+1, \ q+2, \ q+3...$, which are values that match the boundary conditions. The eigenvalues are $\epsilon_{p,q}=(p^2+pq+q^2)\epsilon_{1,0}$, where $\epsilon_{1,0}=4\pi^2/ h^2  $.
    In this work, we focus on a subspace of eigenfunctions, where we set $q=0$, then $p=1,2,3...$ and
\begin{equation}
    \psi_{p,0} (\textbf{r}) = \beta_p \left[ \sin \left(2p\pi y/h\right)-2\sin\left(p\pi y/h\right)\cos \left(p\sqrt{3}\pi x/h\right) \right],
\end{equation}
or
\begin{equation}
    \psi_{p,0}(\textbf{r}) = \beta_p \left[ \sin(pu)+\sin(pv)+\sin(pw)\right],
\end{equation}
where $|\beta_p|=\sqrt{8\sqrt{3}/9L^2}$ is a normalizing constant. The other eigenfunctions can be found in a similar fashion, choosing the appropriate projector (related to the irreducible representations $A_2$ and $E$) to act on the basis function, which can be found in ref. \cite{Li1985}. 

The algebra generators  act as, 
\begin{eqnarray}
    \hat{H} \vert p,0 \rangle & = & \epsilon_{1,0}p^2 \, \vert p,0 \rangle, \\
    \hat{A}^\dagger_{0} \vert p,0 \rangle & = & \sqrt{\epsilon_{1,0}p(p+2)} \, \vert p+1,0\rangle, \\
    \hat{A}_{0} \vert p,0 \rangle & = & \sqrt{\epsilon_{1,0}(p-1)(p+1)} \,  \vert p-1,0 \rangle.
\end{eqnarray}
To determine a position representation for the operators $\hat{A}_0$ and $\hat{A}_0^\dagger$ we first define a projection operator $\mathcal{P}(U)$ that transforms the variables 
\begin{equation}
    \hat{\mathcal{P}}(U): v \rightarrow u , \  w \rightarrow u,
\end{equation}
thus, the vector $(u,v,w)^T$ transforms under the following representation
\begin{equation}
    \hat{\mathcal{P}}(U) =
    \begin{pmatrix}
        1&1&1\\
        0&0&0\\
        0&0&0\\
    \end{pmatrix}
    .
\end{equation}
Therefore, under the action of the projection operator we have 
\begin{equation}
    \hat{\mathcal{P}}(U)\psi_{p,0} = 3\beta_p\sin(pu),
\end{equation}
on the other hand, the action of $\hat{\mathcal{P}}^\dagger(U)$ transform the variables
\begin{equation}
    \hat{\mathcal{P}^\dagger}(U): v \leftarrow u , \  w \leftarrow u,
\end{equation}
and the vector $(u,v,w)^T$ transforms under the representation 
\begin{equation}
    \hat{\mathcal{P}^\dagger}(U) =
    \begin{pmatrix}
        1&0&0\\
        1&0&0\\
        1&0&0\\
    \end{pmatrix}
    .
\end{equation}
In this context, the projector's $ \hat{\mathcal{P}^\dagger}(U)$ action
on an arbitrary function $g(u)$ can be written as
\begin{equation}
    \hat{\mathcal{P}}(U)^\dagger g(u) = g(u)+g(v)+g(w).
\end{equation}
Therefore, one can easily check the recurrence formula
\begin{equation}
    \psi_{p+1,0} = \frac{\beta_{p+1}}{3\beta_p}\hat{\mathcal{P}}^\dagger(U)\left(\cos(u)+\frac{\sin(u)}{p}\frac{d}{du}\right)\hat{\mathcal{P}}(U)\psi_{p,0},
\end{equation}
and setting $\beta_{p+1}/\beta_p=1$ because $|\beta_p|$ does not depends on $p$, it allow us to construct the $\hat{A}_0^\dagger$ in position representation
\begin{equation}
    \hat{A}_0^\dagger = \hat{\mathcal{P}}^\dagger(U)\left(\cos(u)\hat{N}+\sin(u)\frac{d}{du}\right)\hat{\mathcal{P}}(U)\frac{\sqrt{\epsilon_{1,0}\hat{N}(\hat{N}+2)}}{3\hat{N}}.
\end{equation}
It is important to note that the $\hat{A}_0^\dagger$ operator can be written in other forms due to the existent symmetries in this geometry,
\begin{equation}
    \hat{A}_0^\dagger = \hat{\mathcal{P}}^\dagger(V)\left(\cos(v)\hat{N}+\sin(v)\frac{d}{dv}\right)\hat{\mathcal{P}}(V)\frac{\sqrt{\epsilon_{1,0}\hat{N}(\hat{N}+2)}}{3\hat{N}},
\end{equation}
or
\begin{equation}
    \hat{A}_0^\dagger = \hat{\mathcal{P}}^\dagger(W)\left(\cos(w)\hat{N}+\sin(w)\frac{d}{dw}\right)\hat{\mathcal{P}}(W)\frac{\sqrt{\epsilon_{1,0}\hat{N}(\hat{N}+2)}}{3\hat{N}},
\end{equation}
where
\begin{equation}
    \mathcal{P}(V) =
    \begin{pmatrix}
        0&0&0\\
        1&1&1\\
        0&0&0\\
    \end{pmatrix}
    , \qquad { \rm and} \qquad 
    \mathcal{P}(W) =
    \begin{pmatrix}
        0&0&0\\
        0&0&0\\
        1&1&1\\
    \end{pmatrix}
    .
\end{equation}
Henceforth, the annihilation operator is
\begin{equation}
    \hat{A}_0 = \frac{\sqrt{\epsilon_{1,0}\hat{N}(\hat{N}+2)}}{3\hat{N}}\hat{\mathcal{P}}^\dagger(U)\left(\hat{N}\cos(u)-\frac{d}{du}\sin(u)\right)\hat{\mathcal{P}}(U).
\end{equation}

\subsubsection{Coherent States}

\begin{figure}
    \centering
    \includegraphics[width=.8\textwidth]{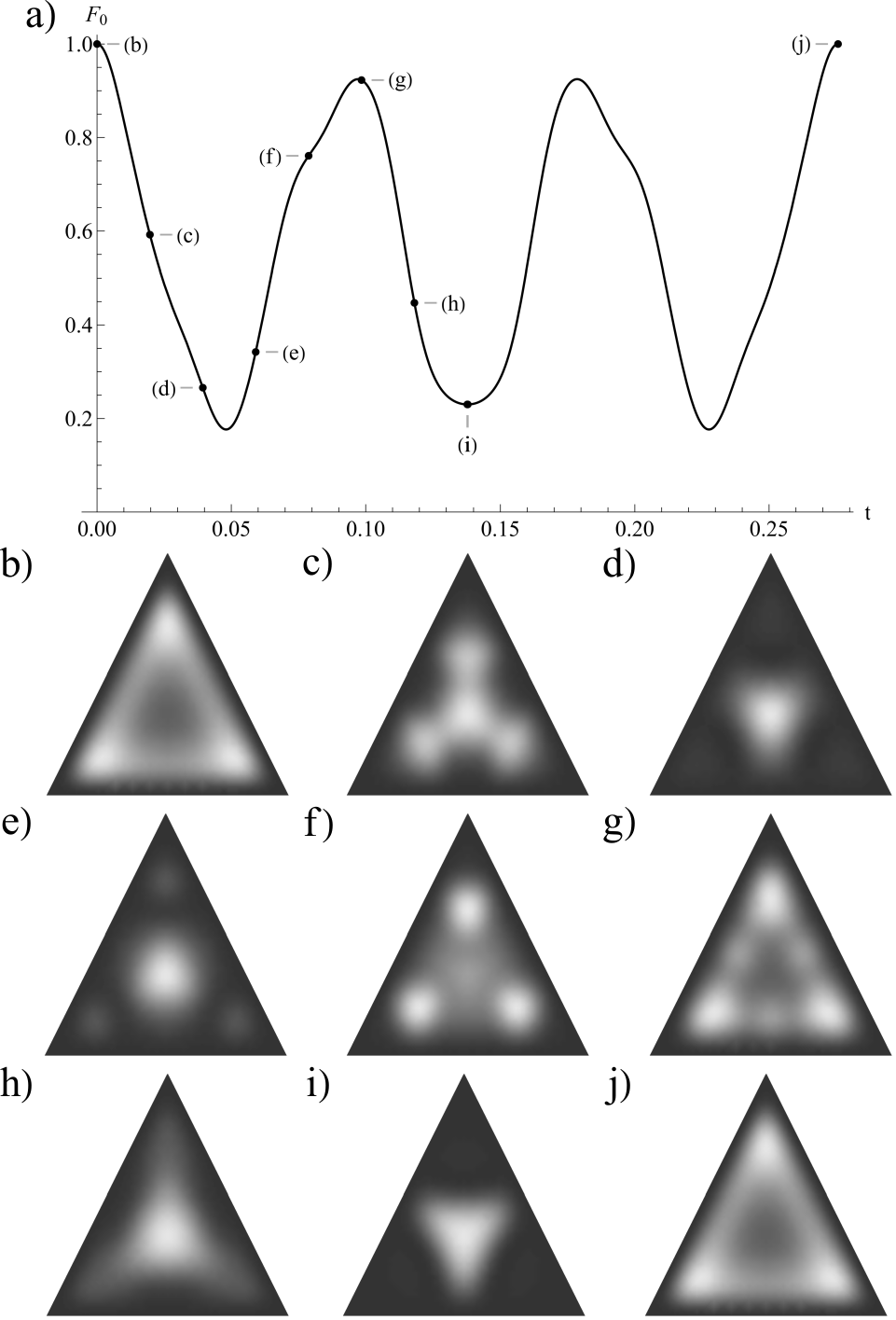}
    \caption{(a) Plot of the fidelity over time in the quantum time revival cycle. (b-i) Density plots of a coherent state $\vert z,t\rangle$ , where $z=5$, in 7 time steps between $t=0$ and $t=T_{rev}/2$. (j) Density plot of the coherent state at the quantum revival time $T_{rev}$. }
    \label{Fig: equilatero}
\end{figure}
The one-dimensional coherent state can be constructed with the denominator
\begin{equation}
    N_{p-1,0}^2! =\frac{\epsilon_{1,0}^{p-1}}{2}(p-1)! \, (p+1)! \, ,
\end{equation}
and the normalizing constant 
\begin{equation}
    N_0(|z|)= \left[2\epsilon_{1,0}I_2(2|z|/\sqrt{\epsilon_{1,0}})\right]^{-1/2} \, ,
\end{equation}
therefore
\begin{equation}
    \vert z \rangle = \frac{1}{\sqrt{I_2\left(2|z|/\sqrt{\epsilon_{1,0}}\right)}}\sum_{p=1}^\infty \left(\frac{z}{\sqrt{\epsilon_{1,0}}}\right)^p\frac{1}{\sqrt{(p-1)! \, (p+1)!}}\vert p,0\rangle \, 
 .
\end{equation}
The time evolution of the coherent state is
\begin{equation}
    \vert z,t \rangle = \frac{1}{\sqrt{I_2\left(2|z|/\sqrt{\epsilon_{1,0}}\right)}}\sum_{p=1}^\infty \left(\frac{z}{\sqrt{\epsilon_{1,0}}}\right)^p\frac{e^{-i \epsilon_{p,0}t }}{\sqrt{(p-1)!(p+1)!}}\vert p,0\rangle,
\end{equation}
and the fidelity $\mathcal{F}_0(t)= \vert\langle z \vert z,t \rangle \vert^2$ is obtained
\begin{equation}
    \mathcal{F}_0(t) = \left\vert\frac{1}{I_2\left(2|z|/\sqrt{\epsilon_{1,0}}\right)}\sum_{p=1}^\infty \left(\frac{|z|}{\sqrt{\epsilon_{1,0}}}\right)^{2p} \frac{e^{-i \epsilon_{p,0}t }}{(p-1)!(p+1)!} \right\vert^2 \, \, .
\end{equation}
In Fig. \ref{Fig: equilatero}a  we plot  the fidelity  over a quantum revival time $t=T_{rev}=3\mu L^2/4\pi \hbar \approx 0.28 $, and perform density plots of the density probability of the coherent state in the position representation

\begin{equation}
        \mathcal{Z}(z,\textbf{r},t) = \frac{1}{\sqrt{I_2\left(2|z|/\sqrt{\epsilon_{1,0}}\right)}}\sum_{p=1}^\infty \left(\frac{z}{\sqrt{\epsilon_{1,0}}}\right)^p\frac{e^{-i \epsilon_{p,0}t }}{\sqrt{(p-1)!(p+1)!}} \psi_{p,0} (\textbf{r}),
\end{equation}
in Fig.\ref{Fig: equilatero}b-j.

\section{Discussion}\label{sec: conclusion}

The generalized Heisenberg algebra provides a framework to develop ladder operators similar to the usual quantum harmonic oscillator for a general one-dimensional quantum system. We developed a generalized Heisenberg algebra (GHA) tailored for quantum billiards. The exploration depicts a subspace of Hamiltonian's eigenfunctions using GHA for any quantum system characterized by quadratic eigenenergies. Key findings include the derivation of closed forms for the normalization constant, the denominator of the coherent state, and the weight function, facilitating the direct application of GHA to obtain coherent states. We find the ladder operators in two representative cases, the square (separable billiard), and the equilateral triangle (non-separable billiard). The formalism is applied to a square billiard, revealing algebra generators and the position representation of ladder operators. It is worth mentioning that the creation of one-dimensional coherent states is demonstrated, followed by an analysis of their time evolution within the quantum revival time. Furthermore, the study extends to the construction of general two-dimensional coherent states and their corresponding time evolution. Having employed two-dimensional coherent states, it was observed that the GHA effectively represented Gaussian-like wave packets moving along a path closely mirroring a classical periodic orbit during a brief time interval. Additionally, we show that the approach can be employed in a non-separable billiard, specifically the equilateral triangle, where algebra generators and one-dimensional coherent states are described. Notably, alternative ladder operators were given due to the symmetries of the equilateral triangle. The position representation of all ladder operators of the GHA is sensitive to the boundary conditions and the choice of the coordinate system. For example, if the square is defined in the region $ (x,y) \in [-L/2,L/2]\times[-L/2,L/2] $ both the eigenfunctions and ladder operators will have different position representations.
The approach contained in this work can be implemented to study other billiards (circular and other triangles). According to ref. \cite{PoschlTellerGHA}, with the machinery of GHA it is possible to obtain positionlike and momentumlike operators and visualize the phase-space. We suggest that this procedure can be useful in studying quantum chaos in billiards by applying it to classically chaotic billiards.

\section*{Acknowledgments}
We thank the financial support from the Brazilian scientific agencies Fundação de Amparo à Pesquisa do Estado do Rio de Janeiro (FAPERJ) and Conselho Nacional de Desenvolvimento Científico e Tecnológico (CNPq), grant related to Programa de Capacitação Institucional (PCI) 2018-2023, number 301098/2024-7.

During the preparation of this work the author(s) used Grammarly in order to improve the article readability. After using this tool/service, the author(s) reviewed and edited the content as needed and take(s) full responsibility for the content of the publication.

% \bibliographystyle{unsrt}
%\bibliography{references}

\end{document}